\documentclass{rmf-d}
\usepackage{nopageno,rmfbib,multicol,times,epsf,amsmath,amssymb,cite}
\usepackage[latin1]{inputenc}
\usepackage[]{caption2}
\usepackage{graphicx}
\usepackage{float}

\def\rmfcornisa{Research OR Education of Physics \hfill\rmf\ {\bf ??} (*?*) ???--???
\hfill MES? A\~NO?}

\def\rmfcintilla{{\it Rev.\ Mex.\ Fis.\/} {\bf ??} (*?*) (????) ???--???}
\clearpage \rmfcaptionstyle 
\begin{document}
\title{   Proton charge radius from a dispersive analysis of the latest space-like $e$-$p$ scattering data 
\vspace{-6pt}}
\author{ Yong-Hui Lin     }
\address{ Helmholtz-Institut~f\"{u}r~Strahlen-~und~Kernphysik~and~Bethe~Center~for
	Theoretical~Physics, Universit\"{a}t~Bonn, \\ D-53115~Bonn,~Germany    }
\author{ }
\address{ }
\author{ }
\address{ }
\author{ }
\address{ }
\author{ }
\address{ }
\author{ }
\address{ }
\maketitle
\recibido{day month year}{day month year
\vspace{-12pt}}
\begin{abstract}
\vspace{1em} 
We present a dispersion theoretical analysis of recent date from electron-proton scattering.
This allows for a high-precision extraction of the electric and magnetic radius of the proton,
$r_E = (0.839\pm 0.002{}^{+0.002}_{-0.003})$~fm and
$r_M = (0.846\pm 0.001{}^{+0.001}_{-0.005})$~fm,
where the first error refers to the statistical type estimated from the bootstrap method, and the second one refers
to the systematic uncertainty related to the underlying spectral functions.    \vspace{1em}
\end{abstract}
\keys{  Proton charge radius, Dispersion theory, Nucleon form factors, $e$-$p$ elastic scattering  \vspace{-4pt}}
\pacs{   \bf{\textit{13.40.Gp,11.55.Fv,14.20.Dh }}    \vspace{-4pt}}
\begin{multicols}{2}

\section{Introduction}

Nucleons and electrons are fundamental blocks that make up everyday matter with
the former accounting for essentially all of its mass. Measurements of nucleon structure
has been one of the most important tasks since the early days of particle physics. The first
measurement on the electric radius, $r_E$, of the proton from the muonic hydrogen, which led to the small radius, $r_E^P= 0.84184(67)\,$fm, with unprecedented precision but differing by 5$\sigma$ from the CODATA value at that time, was reported by Ref.~\cite{Pohl:2010zza} in 2010. 
This glaring discrepancy became well known as the ``proton radius puzzle'' and has ushered a renaissance in the interest in the electromagnetic structure of nucleon in the last decade (see, e.g.,
Refs.~\cite{Carlson:2015jba,Hammer:2019uab,Karr:2020wgh} for recent reviews).

The proton charge radius can be accessed experimentally through the proton electromagnetic form factors which are embedded in the elastic lepton-proton ($ep$ or $\mu p$) scattering but also in the Lamb shift of electronic or muonic hydrogen as performed in Ref.~\cite{Pohl:2010zza}. Recently, new electron-proton scattering data at low momentum transfer were reported by the Jefferson Laboratory (PRad collaboration)~\cite{Xiong:2019umf}, which is the lowest momentum transfer measurement until now (achieved around $10^{-4}~\rm GeV^2$) and has high precision. To extract the proton charge radius, one must parameterize the nucleon form factors with some model and fit it to the scattering data. The parametrization framework inspired by the dispersion theory contains all the physical knowledge we have on the nucleon form factors so far, specifically, it includes all constraints from unitarity, analyticity and crossing symmetry. In addition, it is also consistent with the strictures from perturbative QCD at very large momentum transfer (see Ref.~\cite{Lin:2021umz} for a recent review). In this work, we implement the dispersion theoretical analysis on the latest PRad data together with the precise data from the A1 collaboration at the Mainz Microtron (MAMI)~\cite{Bernauer:2013tpr} and some world data of the nucleon form factors in the space-like region. With a detail investigation on the uncertainties from these experimental data and the underlying formalism, this allows for a high-precision determination of both the electric and the magnetic form factors and the corresponding  charge and magnetic radius, $r_E$ and $r_M$, respectively.

\section{Formalism}

In this section, we collect all necessary theoretical tools, for details see Ref.~\cite{Lin:2021umz}. With the one-photon-exchange assumption, the differential cross section for electron-proton ($ep$) scattering can be expressed through the electric ($G_E$) and magnetic ($G_M $) Sachs form factors as
\begin{equation}\label{eq:xs_ros}
	\frac{d\sigma}{d\Omega} = \left( \frac{d\sigma}
	{d\Omega}\right)_{\rm Mott} \frac{\tau}{\epsilon (1+\tau)}
	\left[G_{M}^{2}(t) + \frac{\epsilon}{\tau} G_{E}^{2}(t)\right]\, ,
\end{equation}
where $\tau = -t/4m_N^2$, with $t$ the four-momentum transfer squared and $m_N$ the nucleon mass, $\epsilon = [1+2(1+\tau)\tan^{2} (\theta/2)]^{-1}$ is the virtual photon polarization. And $\theta$ is the scattering angle of the outgoing electron in the laboratory frame. In addition, $({d\sigma}/{d\Omega})_{\rm Mott}$ is the Mott cross section, which corresponds to scattering off
a point-like proton. In the literature, the nucleon form factors are often displayed as a function of $Q^2$ since $Q^2 \equiv -t >0$ is spacelike in elastic $ep$ scattering. When analyzing the measured cross section data, one usually needs to go beyond the one-photon assumption and consider the two-photon-exchange corrections to Eq.~\eqref{eq:xs_ros}. Here we adopt the same convention as used in Refs.~\cite{Lin:2021umk}. Note that the electric and magnetic radius of the proton are defined as
\begin{equation}
	r_{E/M} = \left(\frac{6}{G_{E/M}(0)}\frac{dG_{E/M}(t)}{dt}\biggr|_{t=0}\right)^{1/2}~.
\end{equation}

For the theoretical analysis, it is convenient to work with the Dirac ($F_1$) and
Pauli ($F_2$) form factors (FFs), which are related to the Sachs FFs by the following linear combinations:
\begin{equation}
	G_E(t) = F_1(t)-\tau F_2(t)~, ~~~~G_M(t) = F_1(t)+F_2(t)~.
\end{equation}
They are normalized at $t=0$, which gives the charge and anomalous magnetic moment of
the proton and the neutron, as
\begin{eqnarray}
	\label{norm}
	F_1^p(0) &=& 1\,, \quad  \; F_1^n(0) = 0\,,\nonumber\\
	F_2^p(0) &=&  \kappa_p\,, \quad  F_2^n(0) = \kappa_n\, ,
\end{eqnarray}
with $\kappa_p=1.793$ and $\kappa_n=-1.913$ in units of the nuclear magneton, $\mu_N = e/(2m_p)$.
When transforming to the isospin basis, the Dirac and
Pauli FFs of proton and neutron will be decomposed into the isoscalar ($s$) and isovector ($v$) parts,
\begin{equation}
	F_i^s = \frac{1}{2} (F_i^p + F_i^n) \, , \quad
	F_i^v = \frac{1}{2} (F_i^p - F_i^n) \, ,
\end{equation}
where $i = 1,2 \,$. 

\begin{figure}[H] 
	\centerline{\includegraphics*[width=0.50\textwidth,angle=0]{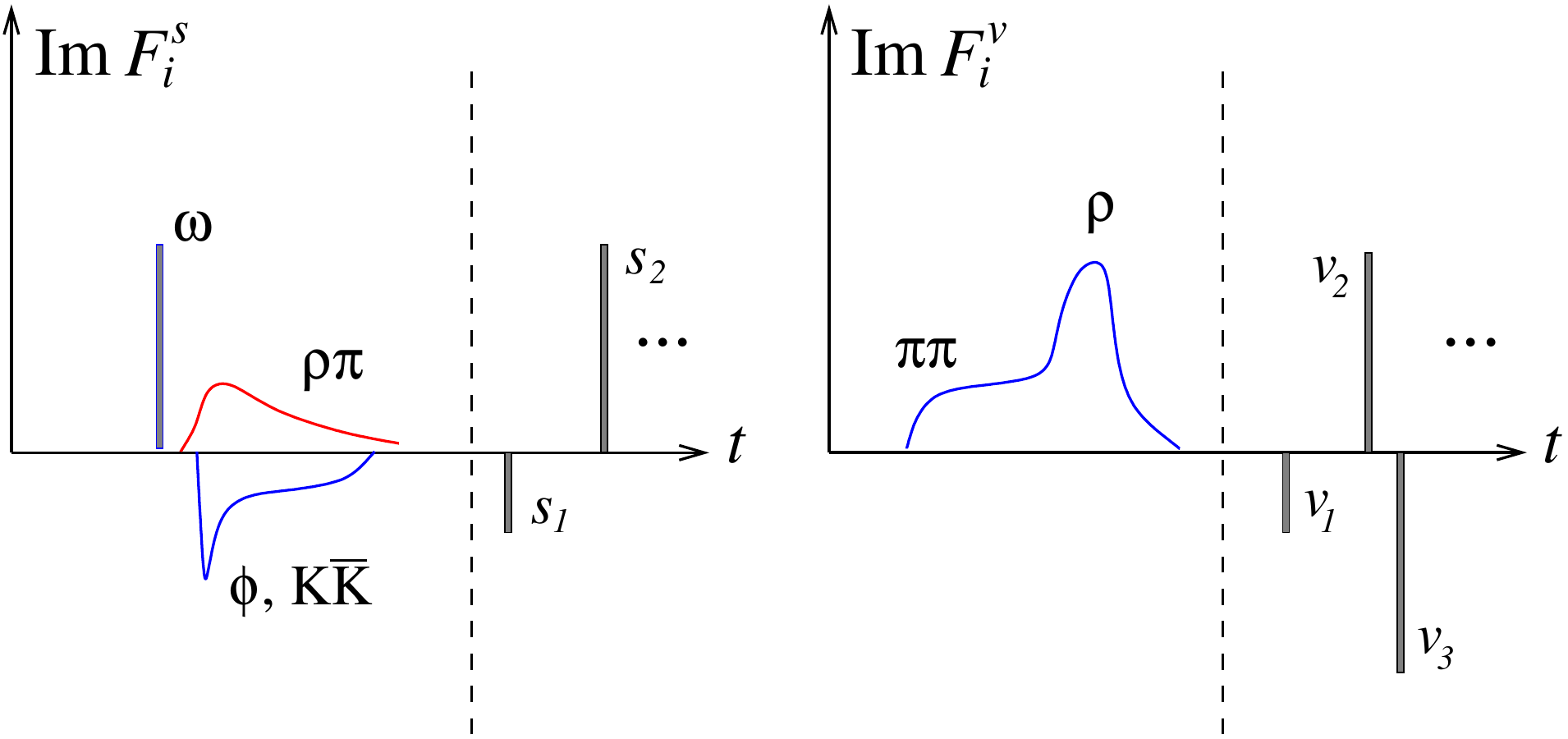}}
	\caption{Cartoon of the isoscalar (left) and isovector (right) spectral function
		in terms of continua and (effective) vector meson poles. The vertical dashed
		line separates the well-constrained low-mass region from
		the high-mass region which is parameterized by effective poles.}
	\label{fig:cartoon}
	\vspace{-3mm}
\end{figure}

The unsubtracted dispersion relations for the nucleon FFs are given by
\begin{equation}
	F_i(t) = \frac{1}{\pi}\int_{t_0}^{\infty}\frac{\text{Im}F_i(t')dt'}{t'-t}~,\hspace{1cm}i = 1,2~,
	\label{disprel}
\end{equation}
where $t_0$ denotes the threshold of the lowest cut of $F(t)$. $t_0 = 4M_\pi^2 \, (9M_\pi^2)$ for the isovector (isoscalar) threshold, with $M_\pi$ the charged pion mass. In our work, the spectral functions are described by means of the spectral decomposition \cite{Chew:1958zjr,Federbush:1958zz} and the vector meson dominance (VMD) model, see Ref.~\cite{Lin:2021umz} for more details. Then the spectral functions can be written as the form:
\begin{align}
	F_i^s (t) &= \sum_{V=\omega, \phi ,s_1,s_2,..} \frac{a_i^V}{m_V^2-t} + + F_i^{\pi\rho}(t) + F_i^{\bar{K}K}(t)~,
	\nonumber\\
	F_i^v (t) &= \sum_{V=v_1,v_2,..}\frac{a_i^V}{m_V^2-t} + F_i^{2\pi}(t)~,
	\label{VMDspec}
\end{align}
with $i = 1,2$. As shown in Eq.~\eqref{VMDspec}, the isoscalar spectral functions contain two lowest poles, $\omega (782)$ and $\phi (1020)$ mesons, and the $\pi\rho$ and
$\bar{K}K$ continua. The isovector spectral functions consist of the two-pion continuum which is estimated from the Roy-Steiner analysis of the pion-nucleon scattering~\cite{Hoferichter:2015hva} and is found to be a critical ingredient for the nucleon form factors as claimed in Ref.~\cite{Hoferichter:2016duk}. And both isoscalar and isovector spectral functions include some effective vector meson poles which contribute to the higher mass parts. A cartoon of the resulting (isoscalar and isovector) spectral functions is shown in Fig.~\ref{fig:cartoon}.

With this dispersion-theoretical parametrization of the nucleon FFs, we can fit to experimental cross sections and nucleon FFs data. Note that the fit parameters in our framework are the various vector meson residua $a_i^V$ and the masses of the effective vector mesons $s_i, v_i$ (the masses of $\omega (782)$ and $\phi (1020)$ are fixed at their physical values). In the fitting procedure, we implement several constraints on those parameters that are refined from the physical knowledge we have about the nucleon FFs. Firstly, we fulfill the normalization conditions as given in Eq.~\eqref{norm}. We also fix the squared neutron charge radius at the recent high-precision determination based on a chiral effective field theory analysis of electron-deuteron scattering~\cite{Filin:2019eoe},
\begin{equation}\label{eq:r2En}
	\langle r_n^2\rangle = -0.105^{+0.005}_{-0.006}~{\rm fm}^2~.
\end{equation}
Moreover, the residua of $\omega (782)$ and $\phi (1020)$) are constrained within the range: 
$0.5\,\mbox{GeV}^2<a_{1}^\omega < 1\,\mbox{GeV}^2$, $|a_{2}^\omega| < 0.5\,\mbox{GeV}^2$
and $|a_{1}^\phi| < 2\,\mbox{GeV}^2$, $|a_{2}^\phi| < 1\,\mbox{GeV}^2$. All other residua are bounded as $|a_i^V| < 5\,$GeV$^2$ due to the naturalness consideration for the couplings. And the masses of the effective poles ($s_1, s_2, \- \ldots , v_1,v_2, \ldots$) are required to be in the range of $1-5$~GeV. 
Finally, the FFs must satisfy the superconvergence relations which are consistent with the requirements of perturbative QCD at very large momentum transfer,
\begin{equation}
	\int_{t_0}^\infty {\rm Im} F_i(t)t^n dt = 0~, ~~i = 1,2~,
\end{equation}  
with $n=0$ for $F_1$ and $n=0,1$ for $F_2$.

Before going to the data fitting, let's briefly introduce our fit strategy. The quality of the fits is measured by means of two different $\chi^2$ functions,
$\chi^2_1$ and $\chi^2_2$, which are defined as
\begin{align}
	\chi^2_1 &= \sum_i\sum_k\frac{(n_k C_i - C(t_i,\theta_i,\vec{p}\,))^2}{(\sigma_i+\nu_i)^2}~,
	\label{eq:chi1}\\
	\chi^2_2 &= \sum_{i,j}\sum_k(n_k C_i - C(t_i,\theta_i,\vec{p}\,))[V^{-1}]_{ij}\notag\\
	& \qquad\qquad \times (n_k C_j - C(t_j,\theta_j,\vec{p}\,))~,
	\label{eq:chi2}
\end{align}
where $C_i$ are the experimental data at the points $t_i,\theta_i$ and
$C(t_i,\theta_i,\vec{p}\,)$ are the theoretical value for a given FF parametrization
for the parameter values contained in $\vec{p}$.
Note that the dependence on $\theta_i$ is applied only to the differential cross sections data. Moreover, the $n_k$ are normalization
coefficients for the various data sets (labeled by the integer $k$ and only used in the fits to
the differential cross section data in the spacelike region), while $\sigma_i$ and $\nu_i$ are 
their statistical and systematical errors, respectively. The covariance matrix is
$V_{ij} = \sigma_i\sigma_j\delta_{ij} + \nu_i\nu_j$. In practice,
$\chi^2_2$ is used for those experimental data where statistical and systematical errors are given separately, otherwise 
$\chi^2_1$ is taken. As done in Ref.~\cite{Lin:2021umk,Lin:2021umz} the various constraints on the form factors are imposed in a soft way. Theoretical errors will be calculated on the one hand using the bootstrap method. On the other hand theoretical errors are estimated by varying the number of effective vector meson poles in the spectral functions. The first error thus gives the uncertainty due to the fitting procedure and the data while the second one reflects the accuracy of the spectral functions underlying the dispersion-theoretical analysis (see Ref.~\cite{Lin:2021umz} for more details).

We are now in the position to analyze the full experimental data. To be specific, for the proton we fit to the cross section data from PRad~\cite{Xiong:2019umf} and from MAMI-C~\cite{Bernauer:2013tpr} as well as to the polarization transfer data on the FFs ratio from Jefferson Lab above $Q^2 = 1\,$GeV$^2$, while only the FFs world data are fitted for the neutron. The size of the data base and the $Q^2$-ranges we are fitting is listed in Tab.~\ref{tab:dbase}. All references for these data can be found in Ref.~\cite{Lin:2021umz}.

\begin{table}[H]
	\vspace{2mm}
	\centering  
	\begin{tabular}{|c|c|c|}
		\hline
		Data type                 &  range of $Q^2$ [GeV$^2$] & \# of data   \\
		\hline
		$\sigma(E,\theta)$, PRad  &  $0.000215 - 0.058$  & 71        \\
		$\sigma(E,\theta)$, MAMI  &  $0.00384  - 0.977$     & 1422      \\
		$\mu_P G_E^p/G_M^p$, JLab  &  $1.18 - 8.49$     & 16        \\
		$G_E^n$, world            &  $0.14 - 1.47$     & 25        \\
		$G_M^n$, world            &  $0.071- 10.0 $     & 23        \\
		\hline
	\end{tabular}
	\caption{Data base used in the fits.}
	\label{tab:dbase}
	\vspace{-3mm}
\end{table}

\section{Results}

As a first test of our parametrization of nucleon FFs, we only fit to the latest PRad data~\cite{Xiong:2019umf}. In this fit, we vary the number of effective vector meson poles from two isoscalar plus two isovector poles (2s+2v) to 5s+5v. Note that 2s+2v means that it contains $\omega$ and $\phi$ meson in isoscalar spectral functions and two additional vector mesons in the isovector spectral functions except those two-body continua that mentioned above. And we only fulfill two normalization related to the proton in Eq.~\eqref{norm} and do not constrain the squared neutron charge radius. It is found that the best fit is given by the configuration of 2s+2v, and all other configurations that contain more poles make the total $\chi^2$ unchanged but the reduced $\chi^2$ increase. $\chi^2/{\rm dof} = 1.33$ for our best fit, completely consistent with the reduced $\chi^2$ obtained in Ref.~\cite{Xiong:2019umf}. The proton radii calculated with that best FFs are given by
\begin{eqnarray}
	\label{eq:rPRad}  
	r_E &=& (0.829\pm 0.012\pm 0.001)\,{\rm fm}~, \notag\\
	r_M &=& (0.843\pm 0.007^{+0.018}_{-0.012})\,{\rm fm}~,
\end{eqnarray}
also consistent with the value, $r_E = (0.831\pm 0.007_{\rm stat}\pm 0.012_{\rm syst})\,$fm, reported by Ref.~\cite{Xiong:2019umf}. Further discussion on the comparison see Ref.~\cite{Lin:2021umk}.

\begin{figure}[H]
	\centering
	\includegraphics[width=0.45\textwidth]{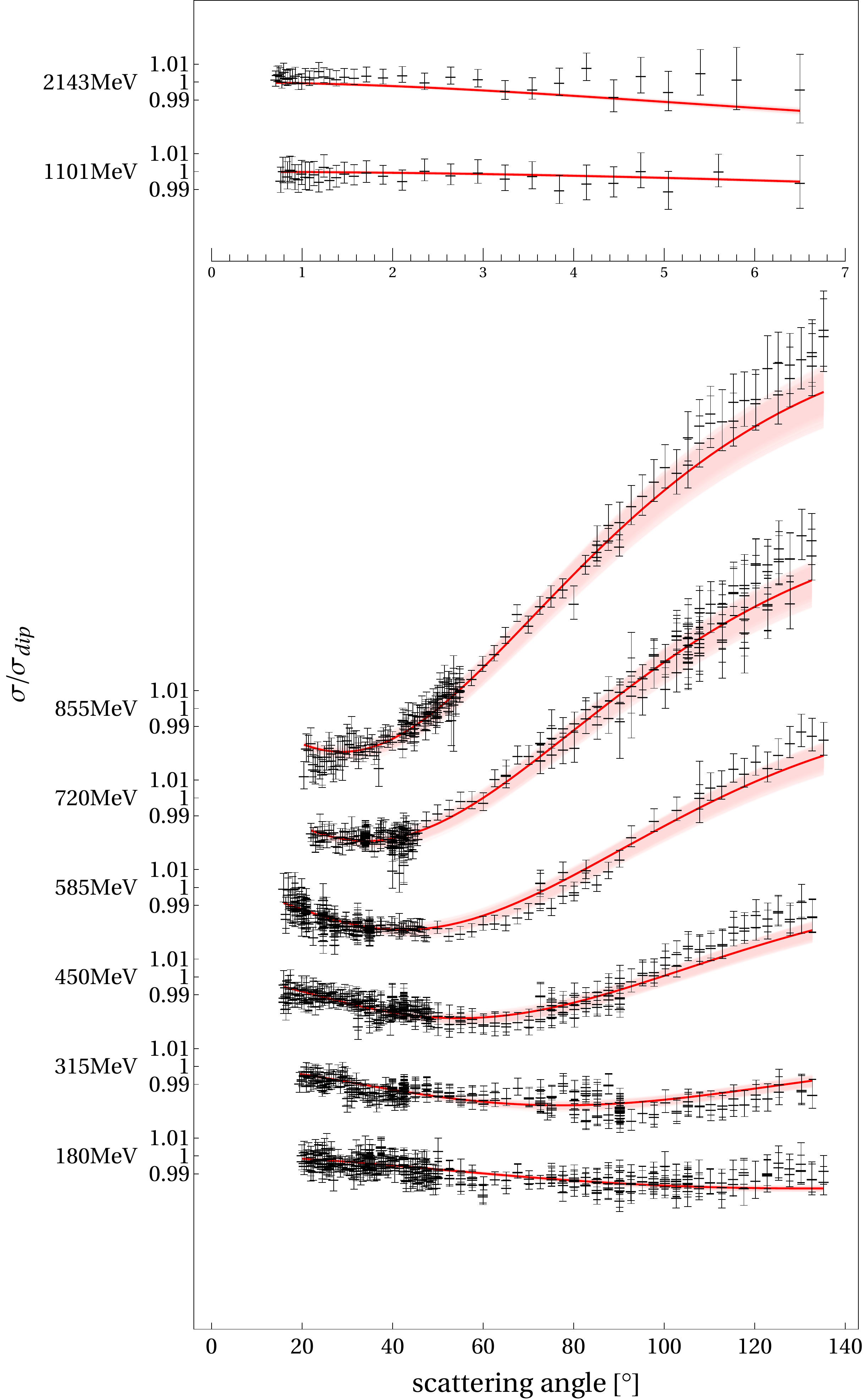}
	\caption{Best fit (solid red line ) to the $ep$ cross section data from PRad (upper panel)
		and MAMI (lower panel) including the two-photon corrections. 
		The red bands give the uncertainty due to the bootstrap procedure. Systematical uncertainties
		are not shown.
		\label{fig:XSbestfit}}
	\vspace{-3mm}
\end{figure}
\begin{figure}[H]
	\centering
	\includegraphics[width=0.45\textwidth]{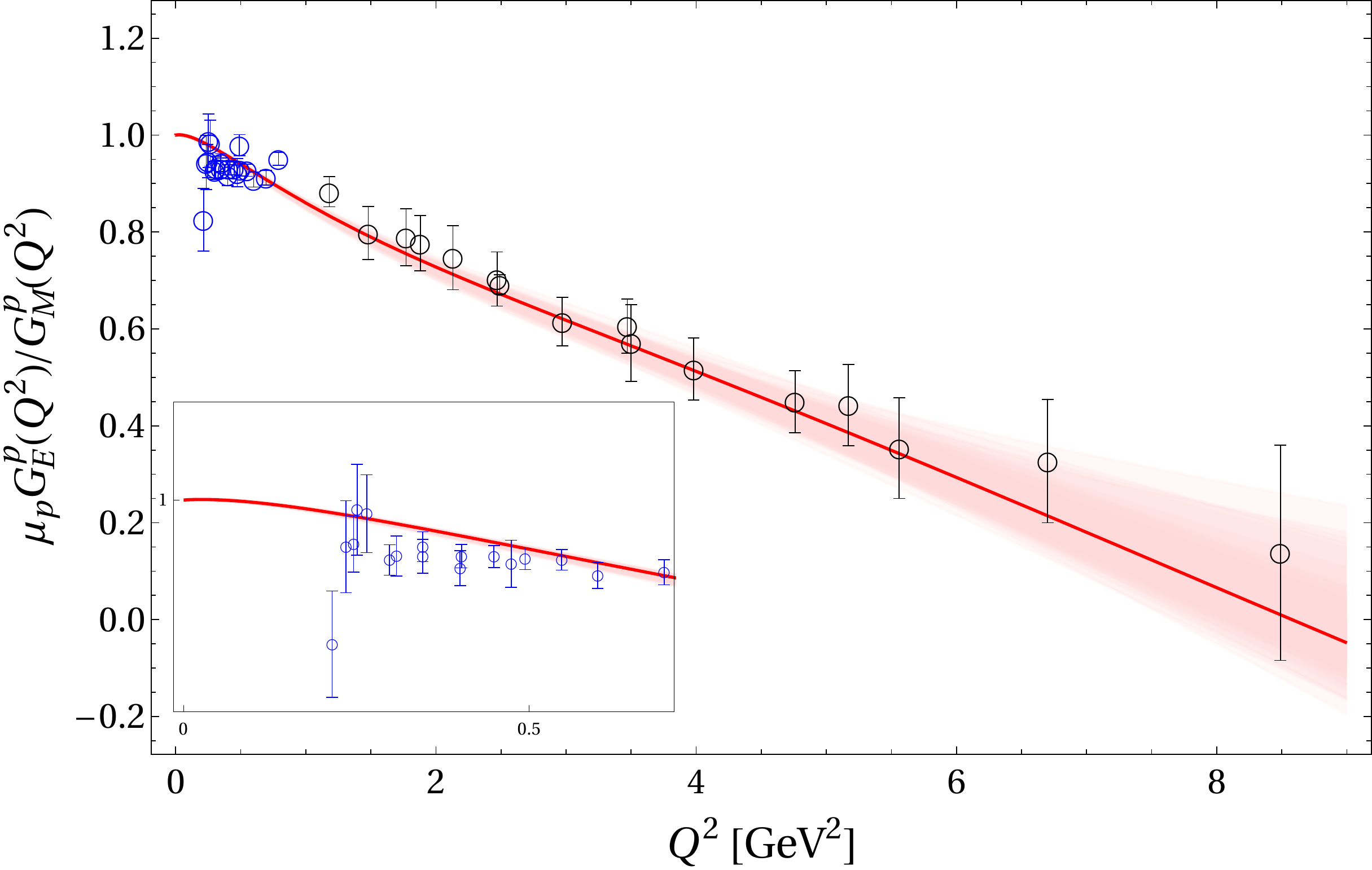}
	\caption{Best fit to the proton form factor ratio data from JLab.
		Note that the blue data (also shown for $Q^2<0.7\,$GeV$^2$ in the inset)
		are not fitted. For notations, see Fig.~\ref{fig:XSbestfit}.
		\label{fig:FFratbestfit}}
\end{figure}
\begin{figure}[H]
	\centering
	\includegraphics[width=0.45\textwidth]{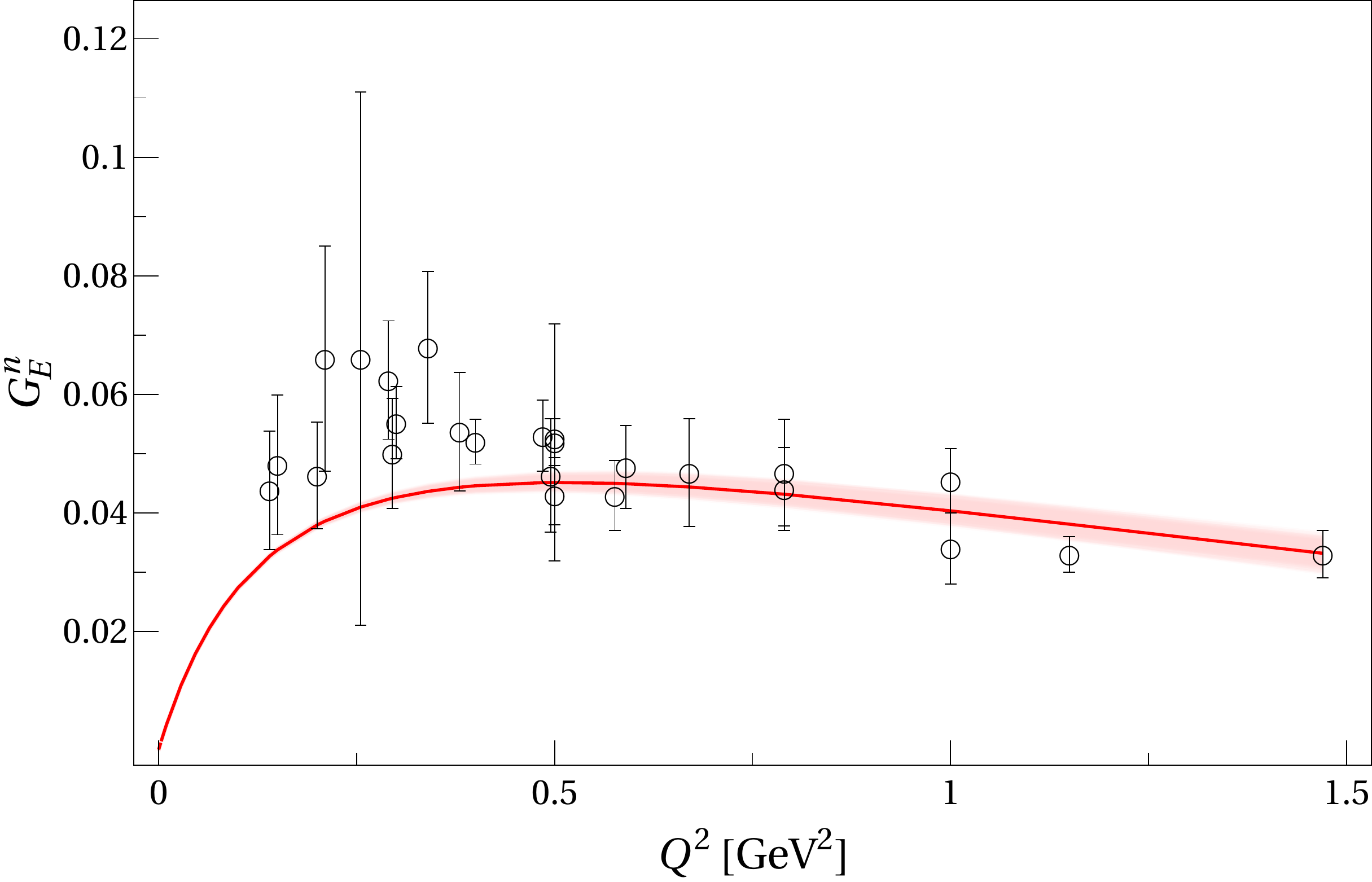}
	\caption{Best fit to the neutron electric form factor data.
		For notations, see Fig.~\ref{fig:XSbestfit}.
		\label{fig:GEnbestfit}}
\end{figure}
\begin{figure}[H]
	\centering
	\includegraphics[width=0.45\textwidth]{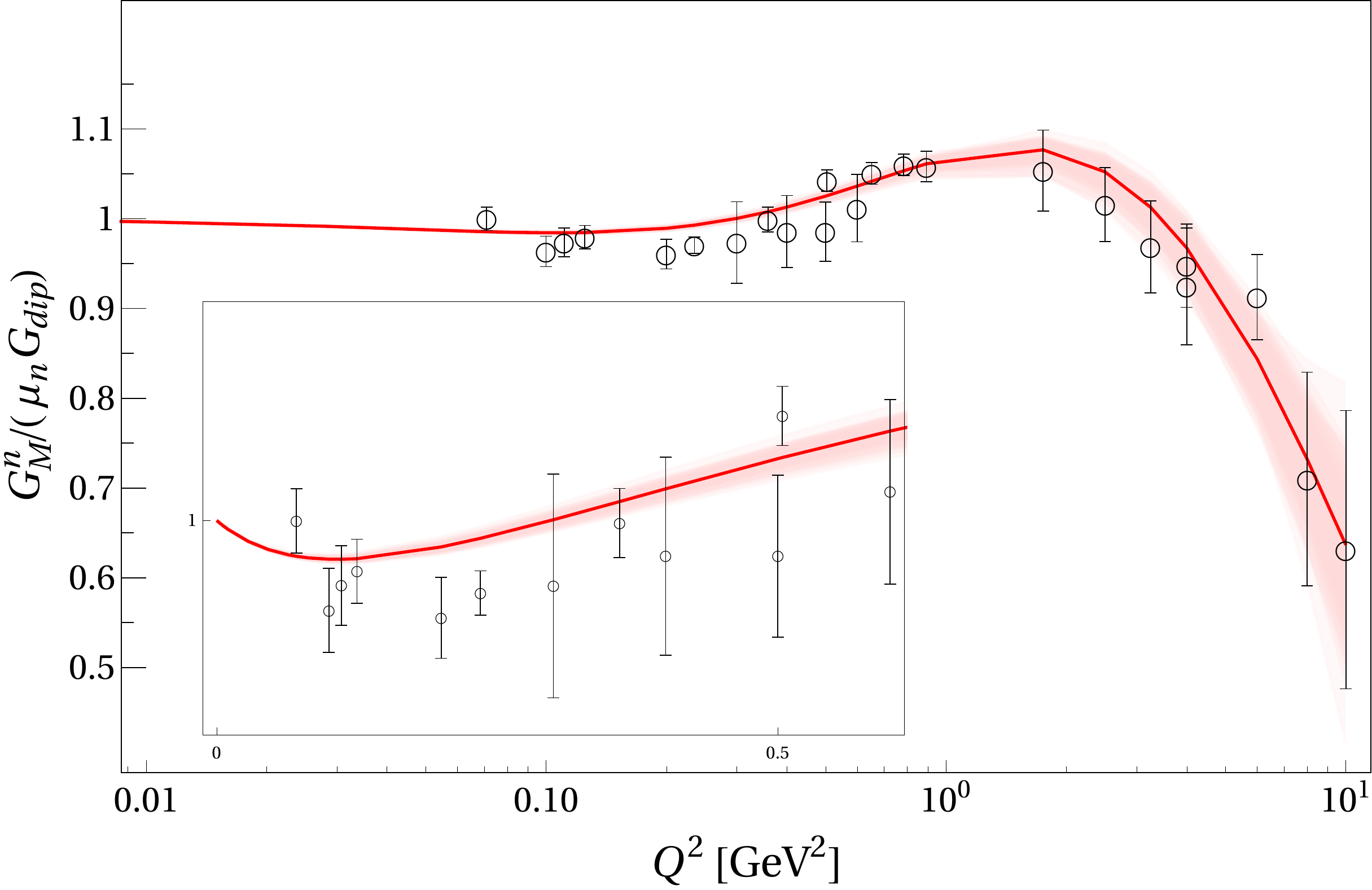}
	\caption{Best fit to the neutron magnetic form factor data.
		For notations, see Fig.~\ref{fig:XSbestfit}.
		\label{fig:GMnbestfit}}
	\vspace{-3mm}
\end{figure}

Next, we move to the combined analysis of all space-like data as listed in Tab.~\ref{tab:dbase}. We search the best fit by varying the configuration of spectral functions from 3s+3v to 8s+8v. The best solution is found to be the 6s+4v configuration where it contains 4 additional effective poles besides $\omega$ and $\phi$ meson in isoscalar sector and 4 effective poles in isovector sector. And all two-body continua are kept in the nucleon FFs. The comparison between our best fit and experimental data is shown in Fig.~\ref{fig:XSbestfit} for the cross section data from PRad and MAMI, Fig.~\ref{fig:FFratbestfit} for the proton FFs ratio from Jefferson Lab (only data above $Q^2 = 1\,$GeV$^2$ are fitted), Fig.~\ref{fig:GEnbestfit} and ~\ref{fig:GMnbestfit} for the neutron electric and magnetic FFs word data together with the error bands estimated from bootstrap sampling, respectively. All these space-like data are described quite well with the error bands of fits and error bar of data considered.

\begin{figure}[H]
	\centering
	\includegraphics[width=0.45\textwidth]{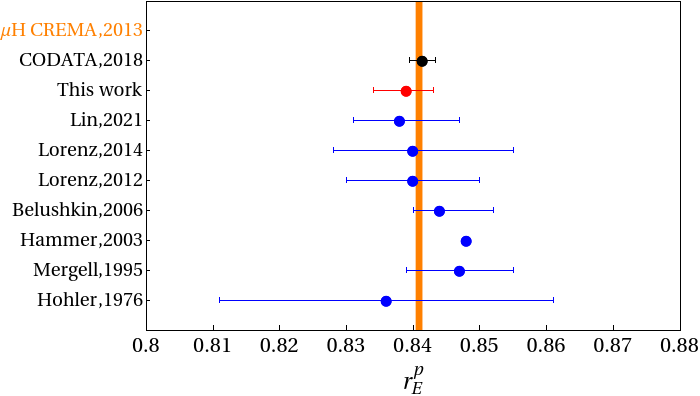}
	\caption{Comparison of the proton radii extracted in this work and other previous dispersion-theoretical extraction. Left y-axis represents the date and author of the corresponding work, see Ref.~\cite{Lin:2021umz} for the relevant papers. The orange band shows the latest radius extraction from the muonic hydrogen~\cite{Antognini:2013txn}.
		\label{fig:rpe}}
	\vspace{-3mm}
\end{figure}

Now, it is time to consider the proton radius. We extract the electric and magnetic radii of the proton from these fits,
\begin{align}
	\label{eq:fullfit}  
	r_E &= (0.839\pm 0.002{}^{+0.002}_{-0.003})\,{\rm fm}~, \notag\\
	r_M &= (0.846\pm 0.001{}^{+0.001}_{-0.005})\,{\rm fm}~,
\end{align}
where the first errors are estimated through the bootstrap procedure and the second ones are obtained by the variation of spectral functions from 3s+3v to 8s+8v. In Fig.~\ref{fig:rpe}, we compare our result with various dispersion-theoretical extractions. Note that here we only list those dispersion-theoretical analyses that include the two-pion continuum explicitly in their spectral functions. What can be clearly seen in this figure is the agreement on the proton charge radius among all theses dispersion-theoretical determinations with the uncertainties considered. And they are in agreement with the value measured from muonic hydrogen~\cite{Antognini:2013txn}. It is shown that from the earliest analysis in 1976 to this work in 2021, the dispersion-theoretical parametrization of nucleon FFs provides a consistent and robust proton charge radius and it gave the small radius even before the muonic hydrogen measurement.   

\begin{figure*}
	\centering
	\includegraphics[width=0.8\textwidth]{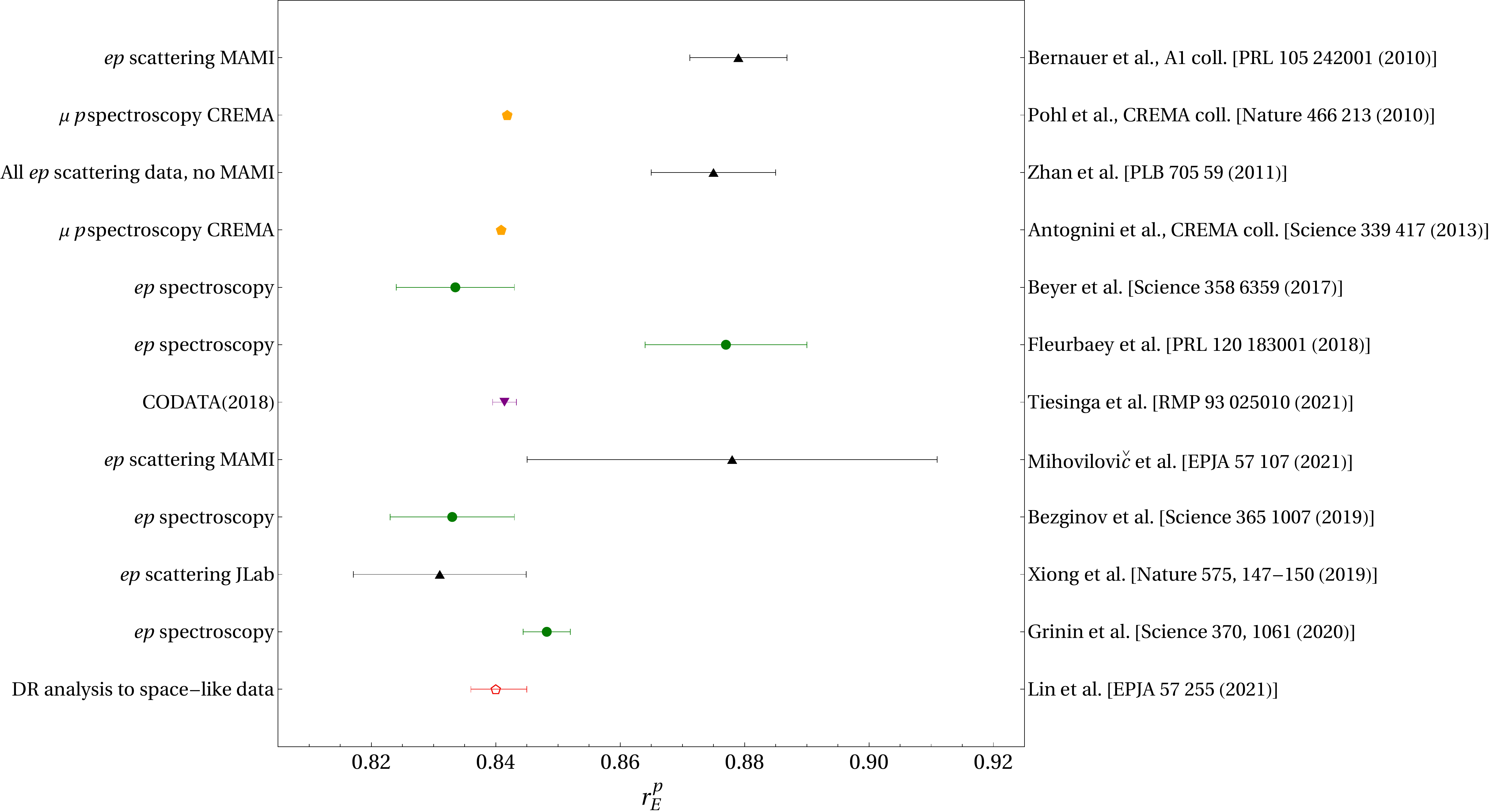}
	\caption{Comparison of the proton radii extracted in this work and other recent papers. Left y-axis represents the process from which the proton radius is extracted and right y-axis shows the corresponding reference.
		\label{fig:rpeoverview}}
	\vspace{-3mm}
\end{figure*}

In Fig.~\ref{fig:rpeoverview}, we compare our determination with recent experimental measurements.
Our results agree quite well with the current CODATA value, $r_E=0.8414(19)$\,fm \cite{CODATAnew} (listed as the purple point). And one can find that both the latest measurements from $ep$ scattering and electronic hydrogen give the small proton charge radius and are consistent with that from muonic hydrogen within the margin of errors. Then one can expect a consistent picture for the proton charge radius appears to emerge as claimed in Ref.~\cite{Hammer:2019uab}.


\section{Summary}

In this work, we have presented the latest dispersion-theoretical analysis of the proton form factors
triggered by the new $ep$ scattering measurement at very low $Q^2$ from the PRad collaboration~\cite{Xiong:2019umf}. With the improved spectral functions worked out in Refs.~\cite{Lin:2021umk,Lin:2021umz}, we have analyzed these new data
as well as the combination of the PRad and other recent space-like data. Together with a detailed investigation on the theoretical uncertainties, we achieved a determination of the proton's electric and magnetic radius with unprecedented precision, as given in Eq.~\eqref{eq:fullfit}. Our results show a consistent value for the proton charge radius with the muonic hydrogen measurements and are also in agreement with various earlier dispersion-theoretical extractions.

\section*{Acknowledgements}
I would like to thank Hans-Werner Hammer and Ulf-G.~Mei{\ss}ner for a most enjoyable collaboration, and specially thank Ulf-G.~Mei{\ss}ner for a careful reading. This work is supported in
part by  the DFG (Project number 196253076 - TRR 110)
and the NSFC (Grant No. 11621131001) through the funds provided
to the Sino-German CRC 110 ``Symmetries and the Emergence of
Structure in QCD",  by the Chinese Academy of Sciences (CAS) through a President's
International Fellowship Initiative (PIFI) (Grant No. 2018DM0034), by the VolkswagenStiftung
(Grant No. 93562), and by the EU Horizon 2020 research and innovation programme, STRONG-2020 project under grant agreement No 824093.

\end{multicols}
\medline
\begin{multicols}{2}

\end{multicols}

\begin{thebibliography}{99}
%
	\bibitem{Pohl:2010zza} 
	R.~Pohl {\it et al.},
	Nature {\bf 466}, 213 (2010).
	
	\bibitem{Carlson:2015jba}
	C.~E.~Carlson,
	Prog. Part. Nucl. Phys. \textbf{82}, 59 (2015)
	[arXiv:1502.05314 [hep-ph]].
	
	\bibitem{Hammer:2019uab}
	H.-W.~Hammer and U.-G.~Mei\ss{}ner,
	Sci. Bull. \textbf{65}, 257 (2020)
	[arXiv:1912.03881 [hep-ph]].
	
	\bibitem{Karr:2020wgh}
	J.~P.~Karr, D.~Marchand and E.~Voutier,
	Nature Rev. Phys. \textbf{2}, 601 (2020).
	
	\bibitem{Xiong:2019umf}
	W.~Xiong, A.~Gasparian, H.~Gao, D.~Dutta, M.~Khandaker, N.~Liyanage, E.~Pasyuk, C.~Peng, X.~Bai and L.~Ye, \textit{et al.}
	Nature \textbf{575}, 147 (2019).
	
	\bibitem{Lin:2021umz}
	Y.~H.~Lin, H.~W.~Hammer and U.-G.~Mei\ss{}ner,
	Eur. Phys. J. A \textbf{57} (2021), 255
	doi:10.1140/epja/s10050-021-00562-0
	[arXiv:2106.06357 [hep-ph]].
	
	\bibitem{Lin:2021umk}
	Y.~H.~Lin, H.~W.~Hammer and U.-G.~Mei\ss{}ner,
	Phys. Lett. B \textbf{816} (2021), 136254
	doi:10.1016/j.physletb.2021.136254
	[arXiv:2102.11642 [hep-ph]].
	
	
	\bibitem{Bernauer:2013tpr}
	J.~C.~Bernauer \textit{et al.} [A1],
	Phys. Rev. C \textbf{90}, 015206 (2014)
	[arXiv:1307.6227 [nucl-ex]].
	
	\bibitem{Chew:1958zjr}
	G.~F.~Chew, R.~Karplus, S.~Gasiorowicz and F.~Zachariasen,
	Phys. Rev. \textbf{110}, no.1, 265 (1958).
	
	\bibitem{Federbush:1958zz}
	P.~Federbush, M.~L.~Goldberger and S.~B.~Treiman,
	Phys. Rev. \textbf{112}, 642-665 (1958).
	
	\bibitem{Hoferichter:2015hva}
	M.~Hoferichter, J.~Ruiz de Elvira, B.~Kubis and U.-G.~Mei\ss{}ner,
	Phys. Rept. \textbf{625}, 1 (2016)
	[arXiv:1510.06039 [hep-ph]].
	
	\bibitem{Hoferichter:2016duk}
	M.~Hoferichter, B.~Kubis, J.~Ruiz de Elvira, H.~W.~Hammer and U.-G.~Mei\ss{}ner,
	Eur. Phys. J. A \textbf{52}, 331 (2016)
	[arXiv:1609.06722 [hep-ph]].
	
	\bibitem{Filin:2019eoe}
	A.~A.~Filin, V.~Baru, E.~Epelbaum, H.~Krebs, D.~M\"oller and P.~Reinert,
	Phys. Rev. Lett. \textbf{124}, 082501 (2020).
	[arXiv:1911.04877 [nucl-th]].
	
	\bibitem{Antognini:2013txn}
	A.~Antognini, F.~Nez, K.~Schuhmann, F.~D.~Amaro, FrancoisBiraben, J.~M.~R.~Cardoso, D.~S.~Covita, A.~Dax, S.~Dhawan and M.~Diepold, \textit{et al.}
	Science \textbf{339} (2013), 417-420
	doi:10.1126/science.1230016
	
	\bibitem{CODATAnew}  
	https://physics.nist.gov/cgi-bin/cuu/Value?rp  
	
\end{thebibliography}
\end{document}